%% file: ms.tex
\documentclass[conference]{IEEEtran}

\usepackage{amsmath}
\usepackage{amssymb}
\usepackage{amsthm}
\usepackage{amsxtra}
\usepackage{amsfonts}
\usepackage{graphicx}
\usepackage{latexsym}
\usepackage{epsfig}
\usepackage{verbatim}
\usepackage{bm}
\usepackage{bbm}
\usepackage{color}
\usepackage{varwidth}
\usepackage{subfigure}
\usepackage{wrapfig}
\usepackage{trivfloat}
\usepackage{todonotes}
\usepackage{url}
\usepackage{fullpage}
\usepackage{dsfont}
\usepackage[margin=9pt,font=small,labelfont=bf,labelsep=colon]{caption}
\usepackage{hyperref}
\usepackage{plain}

\include{randyHeader}
\IEEEoverridecommandlockouts

\title{Robust PCA for Anomaly Detection in Cyber Networks
  \thanks{Approved for Public Release; Distribution Unlimited. Case
    Number 16-4616. \copyright2016 The MITRE Corporation. ALL RIGHTS RESERVED.  Approved 12-23-2016.}}

\date{\today}

\begin{document}

\author{
  Randy~Paffenroth,
  \thanks{R. Paffenroth is with the
    Department of Mathematical Sciences,
    Department of Computer Science, and the Data Science Program,
    Worcester Polytechnic Institute, Worcester, MA 01609,
    email: rcpaffenroth@wpi.edu}
  Kathleen~Kay,
  \thanks{K. Kay is with the
    Department of Mathematical Sciences,
    Worcester Polytechnic Institute, Worcester, MA 01609,
    email: krkay@wpi.edu}
  and Les~Servi
  \thanks{L. Servi is with the MITRE corporation, Bedford, MA 01730,
    email: lservi@mitre.org}
}

\maketitle

\begin{abstract}

This paper uses network packet capture data to demonstrate how Robust Principal
Component Analysis (RPCA) can be used in a new way to detect anomalies which
serve as cyber-network attack indicators.  The approach requires only a few
parameters to be learned using partitioned training data and shows promise of
ameliorating the need for an exhaustive set of examples of different types of
network attacks.  For Lincoln Lab’s DARPA intrusion detection data set, the
method achieves low false-positive rates while maintaining reasonable
true-positive rates on individual packets.  In addition, the method correctly
detected packet streams in which an attack which was not previously
encountered, or trained on, appears.

\end{abstract}

{\bf Keywords:} robust principal component analysis, anomaly
detection, computer networks, cyber defense

\section{Introduction}
\label{introduction}
\input{introduction.tex}

\section{Approach}
\label{approach}
\input{approach.tex}

\section{Problem setup and feature selection}
\label{setup}
\input{setup.tex}

\section{Numerical Results}
\label{results}
\input{results.tex}

\section{Conclusions}
\label{conclusions}
\input{conclusions.tex}

\bibliographystyle{plain}
\bibliography{./bib/CyberWarfare}

\end{document}

%% file: introduction.tex

Over the past decades the dependence of society on interconnected
networks of computers has exponentially increased, with many sectors
of the world economy, such as banking, transportation, and energy,
being dependent on network stability and security.  Accordingly,
maintaining the integrity of computer networks is imperative, and much
research has been performed in this challenging problem domain
\cite{Lakhina2004a,Paffenroth2012a,Kwitt2007,Lakhina2004,Mardani2012}.
In this paper we focus on a sub-topic in this important domain, namely
that of anomaly detection.

Highlights of our results include:

\begin{itemize}
\item We have developed a novel robust principal component approach
  for anomaly detection.  Instead of classic methods where nominal
  background activity (which is presumed to lay on a low-dimensional
  subspace) is defined using some {\it a priori} threshold, we instead
  optimize our thresholds for the current network of interest and
  assume noisy and missing data.
\item As the vast majority of our parameters are trained in an
  unsupervised fashion (i.e., requiring no labeled data), and we only
  have two parameters trained on labeled data, we can make efficient
  use of the limited labeled data available in many real world
  cyber-anomaly detection problems.
\item We demonstrate our performance on three scenarios from the DARPA
  Lincoln Lab Intrusion Detection Evaluation Data Set
  (\url{https://www.ll.mit.edu/ideval/data/2000/LLS_DDOS_1.0.html}). The
  attack scenarios are shown in Table~\ref{scenarios}.
  We trained on the first two scenarios and then, on the third, we
  achieve close to zero false positive rates while maintaining
  reasonable true positive rates even though our algorithm was
  provided no training information for that attack.
\end{itemize}

\begin{table}[h]
\centering
\begin{tabular}{|p{0.5in}|p{2.2in}|}
\hline
  Scenario 1 & IP sweep from a remote site \\ \hline
  Scenario 2 & A probe of live IP addresses looking for a running Sadmind daemon \\ \hline
  Scenario 3 & An exploitation of a Sadmind vulnerability \\ \hline
\end{tabular}
\caption{
  The attack scenarios in the DARPA
  Lincoln Lab Intrusion Detection Evaluation Data Set on which we test our detection
  algorithms.}
\label{scenarios}
\end{table}

\subsection{Background}
Of course, before any progress can be made in the detection of
anomalies, one must carefully consider how such anomalies may be
defined.  In particular, for the automated detection of such anomalies
to be useful for detecting attacks in real-world cyber-data one must
consider their definition both from the computer network perspective
and the mathematical perspective.

An anomaly in a computer network can take many forms.  They include:

\begin{itemize}
\item extreme anomalies, such as a distributed denial of service attack (DDoS),
\item moderate anomalies, such as port scans, and
\item subtle anomalies, such as a buffer overflow attack.
\end{itemize}

\noindent However, for an anomaly to be detectable it must be the case
that it is \emph{different in some way from the normal operations of
  the network}.  For example, if an intrusion detection system (IDS)
is installed on the network, and it performs port scans to detect when
users have inappropriately opened ports, then a port scan is not {\it
  prima faci} an anomaly \emph{for that network}.  One could now
imagine making a long list of rules or templates, such as ``port scans
are anomalies, unless they originate from a specified address'', and
this is precisely how some IDS systems operate.  However, we are
focused on the difficult problem of detecting anomalies where \emph{no
  template for the anomaly is known}.  Accordingly, rather than
enumerating possible anomalies, we focus on \emph{understanding the
  normal operating modes} of a computer network, and define as
anomalous any departure from this normal operating mode.

Accordingly, our goal is to extend the body of literature which aims
to detect anomalies by way of low-dimensional representations of data
measured from computer networks
\cite{Abdelkefi2010,Pascoal2012,Jin2008,Kwitt2007,Mardani2012}.  For
real-world cyber-data, such as network \textbf{p}acket
\textbf{cap}tures (PCAP), it has been observed that the resulting data
often, under normal conditions, resides on a low-dimensional subspace
of the ambient, or measurement, space.  A classic approach to anomaly
detection is to compute the low-dimensional subspace on which the
nominal PCAP data resides and then detect packets that do not lay on
this low-dimensional subspace
\cite{Kwitt2007,Brauckhoff2009,Xu2010a,ringberg2007a}.  Such packets
can be marked as anomalous.

Note, raw packets extracted from PCAP files can sometimes be
difficult to process.  For example, if the packet payloads are
compressed or encrypted, then the dimension of the subspace on which
the packets reside can be unnecessarily large.  Accordingly, in our
work we pre-process the packets to extract features such as port
numbers, IP addresses, packet size, etc.  We have even added
additional derived features, such as whether a packet originated
within the network or is from outside.  As we will demonstrate,
representing packets using such features gives rise to low-dimensional
subspaces and leads to good detection performance.

A classic approach to computing a low-dimensional subspace which
approximates a collection of data is Principal Component Analysis
(PCA) \cite{Eckart1936,HastieTrevorTibshiraniRobertFriedman2009}.
Given a collection of points, PCA computes a linear projection of the
points to a low-dimensional subspace that minimizes the $\ell_2$ error
between the original points and the projected points.

PCA is a workhorse of many data analysis domains including machine
learning and data visualization
\cite{HastieTrevorTibshiraniRobertFriedman2009,Shlens2009a,Abdi2010}.
However, it is well known that the low dimensional subspace provided
by PCA is \emph{sensitive to outliers}.

In particular, outliers will tend to pull the subspace toward the
outlier \emph{quadratically}, making the distance undesirably small
between the computed subspace and the outliers one wishes to detect.
More precisely, PCA computes a family of low-dimensional subspaces,
and the user is required to select which $k$-dimensional subspace they
believe is the best representation of the data.  However, this
selection is made difficult when each subspace is computed from a
mixture of nominal and anomalous measurements.  As our aim is anomaly
detection in real-world network data, a more delicate analysis is
likely required.

\subsection{Potential of RPCA}

Accordingly, herein our focus is on the growing field of \emph{robust}
principal component analysis (RPCA).  RPCA has a large and active
extent literature \cite{Candes2009, Chandrasekaran2009, Candes2011,
  Paffenroth2012a, Paffenroth2013b,Wright2009}, and there are many
algorithms that focus on the \emph{recovery} of low-dimensional
subspaces from data which has been corrupted by outliers.

In particular, much work has been performed on \emph{recovery}
problems where theorems are proved, and numerical demonstrations
provided, along the lines of:

\begin{quotation}
  \noindent Assuming there exists a true low dimensional subspace $L$ and a true
  collection of anomalies $S$, one \emph{recovers} approximations of
  these values from their sum $M=L+S$ (or some other similar
  combination of $L$ and $S$) assuming $L$ and $S$ satisfy some
  conditions \cite{Candes2011,Wright2009}.
\end{quotation}

\noindent Such problems are quite non-trivial and have
received much attention \cite{Candes2011,Wright2009}.  In particular, since many
combinations of $L$ and $S$ give rise to the same $M$, recovering a
specific desired pair $L, S$ purely from such an $M$ often requires
delicate analysis.

However, as opposed to such \emph{recovery} problems, herein we take a
novel approach and instead concern ourselves with \emph{detection}
problems.  Hearkening back to the case of classic PCA, one can ask two
quite different questions:

\begin{enumerate}
\item Given some data (perhaps corrupted by noise) can one
  \emph{recover} the true low dimensional subspace that spans the
  data (with the corruptions removed)?
\item Given some data can one find a low dimensional subspace that is
  most effective for some other downstream \emph{detection} algorithm.
  I.e., how is PCA best used as a preprocessing step for some other
  machine learning algorithm?
\end{enumerate}

\noindent Both approaches are used quite widely in the PCA literature
\cite{ringberg2007a} and it is questions similar to the second question above
that inspire our approach here.  For example, when using PCA as a
preprocessing step for some downstream algorithm, it is quite classic
to consider either a target dimension $k$ for the computed subspace
or, equivalently, a desired threshold $\gamma$ for the singular values
of a low-rank data matrix $L$.  Such a threshold is equivalent to a
statement about the maximum acceptable error between the original data
points and their projection.

Accordingly, $\gamma$ can either be chosen based upon some
\textit{a priori} target error tolerance in the singular values or, as
we propose here, \emph{chosen based upon the cross validated detection
  performance of the detection algorithm}.  Similarly, one can ask two
different questions of RPCA.  As we will discuss in detail in the
sequel, the RPCA algorithms that we leverage herein have a parameter
$\lambda$ which controls the trade-off between the low-rank matrix $L$
and the sparse matrix $S$.  Mirroring the idea in the PCA case, we ask
the following question:

\begin{enumerate}
\item Given some data (perhaps corrupted by noise), how should one
  $\lambda$ that best leads to the \emph{recovery} of the true
  low-dimensional subspace and the true sparse anomalies from which
  the observed data was constructed?
\item Given some data, how should one find a dimension $k$ \emph{and a
    value of $\lambda$} that lead to a low-dimensional subspace and a
  set of anomalies that is most effective for some other downstream
  \emph{detection} algorithm that only gets to train $\lambda$ and not
  the numerous other parameters of the algorithm?  I.e., how is RPCA
  best used as a preprocessing step for some other machine learning
  algorithm?
\end{enumerate}

\noindent In one of the key novelties of our analysis, we demonstrate
that a cross validated choice of $\lambda$ rather than choosing
$\lambda$ based upon some recovery principle, can lead to
substantially improved algorithms for detecting anomalies in computer
networks.  In particular, there are no current approaches that use
RPCA for detecting anomalies in computer networks, of which we are
aware, that avail themselves of the full flexibility of RPCA.  In
particular, they do not select the key parameter $\lambda$ based upon
a cross validation principle.

In particular, herein we demonstrate the efficacy of our approach on
PCAP measurements from the Lincoln Labs DARPA Intrusion Detection Data
Set (\url{https://www.ll.mit.edu/ideval/data/2000/LLS_DDOS_1.0.html}),
which cover a swath of normal computer network operations and a
variety of different attack scenarios.  Perhaps most interestingly,
using our methods we are able to train our parameters such as
$\lambda$ on a small set of attacks, but then we are able to use these
same parameter settings to detect different attack modalities \emph{on
  which the algorithm was not trained}.

In Section~\ref{approach} we describe the mathematical foundation of
our approach, in Section~\ref{setup} we describe the setup of our
experiments and the data set we use for validation, in
Section~\ref{results} we demonstrate our results, and in
Section~\ref{conclusions} we provide a summary and pointers to future
work.  Finally, in the Appendix we provide additional notes on
the effecient solution of the problems of interest using the Augmented
Lagrangian Method and the Alternating Direction Method of Multipliers.


\subsection{Contribution and previous work}


PCA is a standard algorithm in many problem domains
\cite{HastieTrevorTibshiraniRobertFriedman2009} and it has been widely
applied in computer network analysis \cite{Brauckhoff2009} (and
reference therein) with seminal work going back to at least 2004
\cite{Lakhina2004a}.  In addition, there are several examples of RPCA
being used in the extant literature for computer network anomaly
detection
\cite{Abdelkefi2010,Pascoal2012,Jin2008,Kwitt2007,Mardani2012} (and
references therein) with a quite recent review to be found in the 2015
Ph.D. thesis \cite{Mardani2015} and related paper \cite{Mardani2012}.
However, only the more recent references
\cite{Pascoal2012,Abdelkefi2010,Mardani2012} use the modern convex
nuclear norm approaches we leverage here.

In particular, \cite{Abdelkefi2010} uses a RPCA on data which is
similar to our own, but they take the opposite approach to ours when
considering the coupling constant $\lambda$.  They use the theoretical
value suggested in \cite{Candes2011} and do not study the interplay
between $\lambda$ and the performance of the downstream detection
algorithms.

Accordingly, a key novelty of our approach is a careful treatment of the
parameter $\lambda$ which is the key element in balancing the importance of
low-dimensional $L$ and the anomalous $S$.  In particular, by studying the
interplay between $\lambda$ and downstream detection algorithms, the quality of
our detection results are greatly enhanced as compared to PCA.  In addition, we
analyze a detection threshold on the anomalies $S$, here denoted by $\alpha$,
which in our experiments turns out to essentially be $1$ given a judicious
choice of $\lambda$.


%% file: approach.tex
\subsection{PCA}
\label{PCAsection}

We begin the derivation of our methodology by considering a collection of
points $\{y_0,...,y_n\}$ with $y_i \in \mathbb{R}^m$.  Each point can
be thought of a collection of features that represent a measurement of
our computer network.  For example, as we will detail in
Section~\ref{setup}, herein we view each $y_i$ as a collection of
features derived from a single Internet packet (such as IP address,
port number, etc.).  Therefore, the collection of points
$\{y_0,...,y_n\}$ represents some collection of packets measured
across different computers and times.  Given such a collection of
points $\{y_0,...,y_n\}$, the goal of PCA is to compute a linear
projection of the points $\{\hat{y}_0,...,\hat{y}_n\}$ with each of
the $\hat{y}_i$ laying on a specified $k$-dimensional subspace of
$\mathbb{R}^{n}$.\footnote{Note, there are two equivalent meanings
  of the phrase ``laying on a specified $k$-dimensional subspace''.
  First, one can consider that $\hat{y}_i \in \mathbb{R}^m$ and that
  the $m$ coordinates of each $\hat{y}_i$ are linear combinations of
  the $k$ vector that span the subspace.  Second, one can consider
  that $\hat{y}_i \in \mathbb{R}^k$ and that the $k$ coordinates are
  the position of the point on the $k$ dimensional subspace itself.}

To specify the desired $k$-dimensional subspace of $\mathbb{R}^{n}$
one can encode the original points $y_i$ into a data matrix
$[y_0,...,y_n] = Y \in \mathbb{R}^{m \times n}$ (i.e., by having each
$y_i$ be a column of $Y$).  One can then compute the desired
$k$-dimensional subspace by solving the optimization problem

\begin{align} \label{PCAopt}
  &\min_{L} \| L-Y \|_F^2 \\ \nonumber
  \text{subject to}\quad & \rho(L) \le k
\end{align}
\noindent where $L \in \mathbb{R}^{m \times n}$, $\| L-Y \|_F^2$ is
the Frobenius norm of $L-Y$ (i.e., the sum of the squares of the
entries of $L-Y$), and $\rho(L)$ is the \emph{rank} of $L$
(i.e., $\rho(L)$ is the number of non-zero singular values of the matrix
$L$).

A more common derivation of PCA is in terms of the Singular Value
Decomposition (SVD) \cite{Eckart1936} rather than as an optimization as in
\eqref{PCAopt}.  However, the optimization point of view will have an
important role to play in the sequel.

In particular, given a matrix
$Y \in \mathbb{R}^{m \times n}$, one can always write

\begin{eqnarray}
  Y = U \Sigma V^T
\end{eqnarray}

\noindent where $U \in \mathbb{R}^{m \times m}$ and
$V \in \mathbb{R}^{n \times n}$ are unitary (i.e., $U U^T = I$ and
$V V^T=I)$, and $\Sigma$ is diagonal.  In addition, the diagonal
entries of $\Sigma$, denoted $\Sigma_{ii} = \sigma_i$ are called the
singular values of $Y$.

\begin{figure}[htbp]
  \centering
  \includegraphics[width=0.45\textwidth]{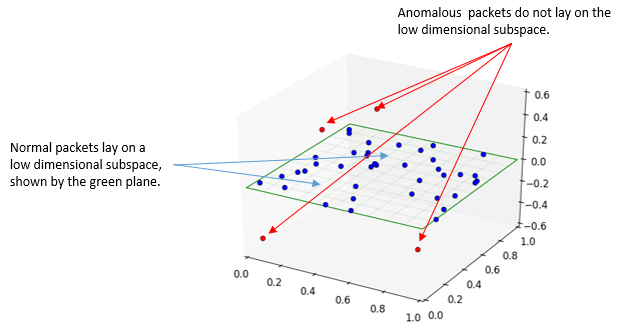}
  \caption[Caption.]{In this figure we show an example of the key idea
    of anomaly detection by way of PCA.  Nominal data is assumed to
    lay on a low-dimensional linear subspace and anomalies are defined
    as any departures from this low-dimensional subspace.}
\label{experiment1}
\end{figure}

In a seminal result, Eckart and Young in their 1936 paper \cite{Eckart1936}
proved that the optimization problem in \eqref{PCAopt} can be solved
in closed from by setting

\begin{eqnarray}
  L = U \hat{\Sigma} V^T
\end{eqnarray}

\noindent where $\hat{\Sigma}$ is computed from $\Sigma$ by setting the $m-k$
smallest singular values $\sigma_i$ to $0$ (i.e., by retaining only the $k$
largest singular values), and $L$ is therefore a low-rank approximation of $Y$.
Similarly, one can compute a projection of $Y$ onto the coordinate system of the
$k$-dimensional subspace by using the formula

\begin{eqnarray}
  X = \hat{\Sigma} V^T
\end{eqnarray}

\noindent and removing the zero rows of $X$ arising from the $0$
values on the diagonal of $\hat{\Sigma}$.  An example of PCA is
shown in Figure~\ref{experiment1}.
Note, the relationship between the optimization view of PCA, as shown
in \eqref{PCAopt}, and the linear algebra view, as exemplified by the
SVD, will be important in the sequel.  In particular, while the SVD is
the most common implementation of PCA in current use, it is actually
the optimization version that inspires much recent work.  In
particular, there are many recent results in robust versions of PCA
methods which revolve around hearkening back to the optimization roots
of PCA.

\subsection{RPCA}

\begin{figure}[htbp]
  \centering
  \includegraphics[width=0.45\textwidth]{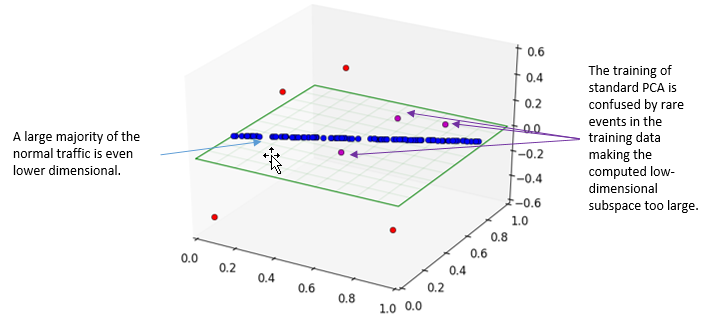}
  \caption[Caption.]{
    In this figure we show an example of the key idea of the RPCA algorithm.
    The nominal data is corrupted by rare events (either anomalous or not) and
    this data makes the computed dimension too large.  The truly nominal data is
    lower dimensional than PCA reveals.  I.e., even with a large threshold for
    the singular values, PCA would conclude that the data is two-dimensional.
    RPCA detects these events and provides a better representation of the
    low-dimensional state of the network.}
 \label{experiment2}
\end{figure}

PCA, while a widely used technique for dimensionality reduction, is
unfortunately sensitive to the presence of outliers.  In particular,
as shown in Figure~\ref{experiment2}, even when the vast majority of
the measurements in $Y$ lay on a low-dimension subspace, the presence
of just a few outliers can substantially increase the dimension of the
subspace produced by the PCA algorithm and lead to a reduction in our
ability to later detect anomalies.  Note, this is not an issue that
can be fixed by merely a judicious choice of $k$.  Every singular
value produced by \eqref{PCAopt} can, and likely would be, perturbed
by even a single outlier.  In particular, outliers can easily
transform singular values which, for purely nominal data, would be
close to $0$ to arbitrarily large values, and thereby increase the
detected dimension.

Fortunately, there has been a flurry of recent attention paid to
\emph{Robust} Principal Component Analysis (RPCA) and excellent
progress has been made in the literature.  We provide a very brief
overview of the main ideas, and refer the reader to \cite{Candes2009,
 Candes2011, Chandrasekaran2009, Cai2010} for more
details.  In this section we closely follow the notation and
derivation from \cite{Paffenroth2012a,Paffenroth2013b}

As noted in \cite{Candes2011}, the RPCA problem may seem daunting upon
preliminary inspection.  Given a measurement matrix $Y$, we must tease
apart a low rank matrix $L$ and a set of sparse anomalies $S$
\textit{without knowing \textit{a priori} the true dimension $k$ of
  $L$, nor knowing the number or locations of the anomalous entries in
  $S$.}  Similar to \eqref{PCAopt}, this problem can be phrased as an
optimization problem by writing

\begin{align} \label{npalg}
  &\min_{L,S}\rho(L)+\lambda\|S\|_{0}\\ \nonumber
  \qquad \text{subject to} \quad &
                                   |Y-(L+S)| = 0
\end{align}

\noindent where $\rho(L)$ is the rank of $L$, $\|S\|_{0}$ is the
number of non-zero entries in $S$, and $\lambda$ is a coupling
constant which controls the trade-off between the low-rank matrix $L$
and the sparse matrix $S$.  Unfortunately, as opposed to
\eqref{PCAopt}, we do not have any closed form solution to
\eqref{npalg}.   Even worse, a n\"{a}ive, brute force
approach to the problem, where one searches over all possible
combinations of low-rank matrices $L$ and entries of $S$ corresponding
to a presupposed number of anomalies, would be NP-hard in the number
of anomalies.

However, Theorem 1.2 in \cite{Candes2011}, Theorem 2.1
\cite{Paffenroth2012a}, and many similar theorems in the extent
literature provide remarkable guarantees for recovery $L$ and $S$.
Providing details for these theorems would take us too far afield in
the current context, and the interested reader may refer to extent
literature for details \cite{Candes2009, candes09ex, Chandrasekaran2009,
  Candes2011, Paffenroth2012a, Paffenroth2013b}.
Herein we merely observe that the optimization in \eqref{npalg} is
NP-hard, but a closely related problem can be solved if some technical
conditions are met.\footnote{Classically, these conditions bound the
  rank of $L$, bound the sparsity of $S$, require that the columns of
  $L$ are \textit{incoherent} far from the standard basis, and require
  that the non-zero entries in $S$ are distributed uniformly.}

In particular, assuming such conditions are met, then, with high
probability the \emph{convex} program

\begin{align} \label{mainalg}
  &\min_{L,S}\|L\|_{*}+\lambda\|S\|_{1}\\ \nonumber
  \qquad \text{subject to} \quad &
                                   |Y-(L+S)|
                                   \preceq \epsilon
\end{align}

\noindent recovers $L$ and $S$, where
$\|L\|_{*} = \sum_{i=1}^m\sigma_{i}$ is the nuclear norm of $L$ (i.e.,
the sum of the singular values of $L$) and
$\|S\|_1:= \sum_{ij}|S_{ij}|$.  $\lambda$ is as in \eqref{npalg} and
$\epsilon$ is a set of point-wise error constraints which we used to
ameliorate the noise found in real-world data.  The reader familiar
with such algorithms will no-doubt note that $\|S\|_1$ is a convex
relaxation of $\|S\|_0$, and $\|L\|_*$ is a convex relaxation of
$\rho(L)$, and such problems can be efficiently solved
\cite{Boyd2010a, Candes2009, Candes2011, Paffenroth2012a,
  Paffenroth2013b, Halko2011}.

Note, in \eqref{mainalg}, the importance of the parameters $\lambda$
and $\epsilon$.  In particular, Theorem 1.2 in \cite{Candes2011}
proves that setting $\lambda = \frac{1}{\sqrt{\max(m,n)}}$, where
$Y \in \mathbb{R}^{m \times n}$ guarantees the \emph{recovery} of $L$
and $S$ from $Y$ (assuming the constraints mentioned previously).

\emph{However, in the current context, the recovery of a presumed $L$
  and $S$ is not our goal}.  In particular, as we are attempting to
detect anomalies in real-world measured data, it is not clear what
such a ``true'' $L$ and $S$ would mean, even if we were to compute
them.  Accordingly, in our work, we view $\lambda$ and $\epsilon$ as
parameters to be estimated from training data, and tuned to our
particular detection task.  Of course, having $\lambda$ and $\epsilon$
as parameters we learn from data requires us to be in possession of
appropriate training data.  However, as we will demonstrate in
Section~\ref{results}, it is our view that the additional requirements
for training data are well worth the substantially better performance
we achieve.

\begin{figure}[htbp]
  \centering
  \includegraphics[width=0.45\textwidth]{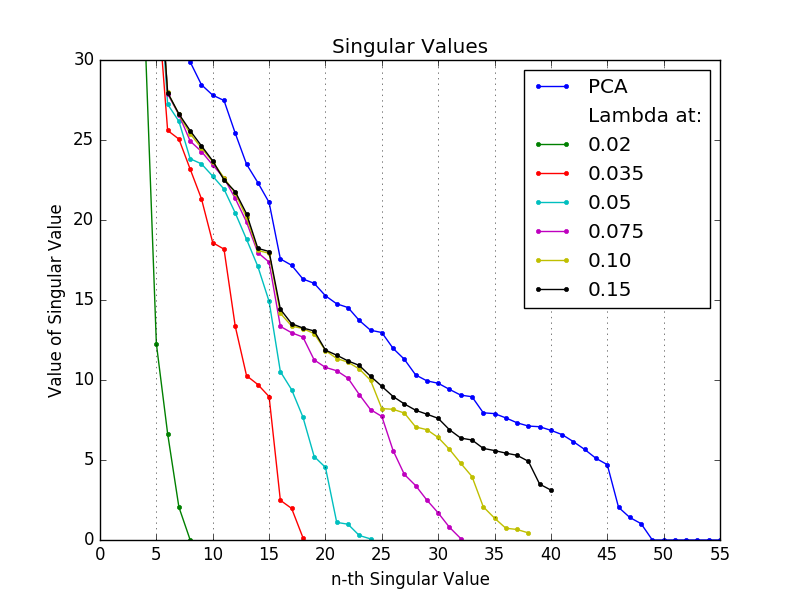}
  \caption[Caption.]{This figure shows the number of non-zero singular values
  	of various low-rank matrices $L$ computed from the same measurement matrix
  	$Y$. These $L$-matrices are calculated with values of $\lambda$ ranging
  	from 0.02 through 0.15, whose lines are labeled as such. The PCA line
  	is equivalent to sending $\lambda$ to infinity. As $\lambda$ increases,
  	the number of non-zero singular values increases to our maximum allowed
  	rank (40). $\lambda = 0.15$ has 40 non-zero singular values (and does not
  	reach our iteration threshold (1000)). The SVD of $Y$ has 94 singular values
  	but the latter 55 are the same. In calculation, we will be treating this
  	as 49 non-zero singular values. As for $\lambda = $ 0.02, 0.035, 0.05,
  	0.075, and 0.10, we have 8, 18, 24, 32, and 38 non-zero singular values,
  	respectively.
    Larger values of $\lambda$ makes placing anomalies into $S$ more
    expensive, and generally increase the rank. On the other hand, smaller
    values of $\lambda$ makes placing anomalies into $S$ less expensive. Interestingly,
    the relation between $\lambda$ and the number of non-zero singular values
    is not necessarily linear.
    }
 \label{singularvalues}
\end{figure}

As a foreshadowing of those results, see
Figure~\ref{singularvalues}. There we show the singular values of $L$
computed from a collection of PCAP features encoded in a measurement
matrix $Y$ (which we will detail in the next section).  Observe how
various choices of $\lambda$ have a \emph{profound} effect on the
dimension of the computed low-rank matrix $L$.  As one might imagine,
the ability to choose an appropriate value of $\lambda$, and therefore
an appropriate low-rank $L$, has an equally profound effect on the
ability of the algorithm to detect anomalies.

%% file: setup.tex
\subsection{Problem setup}
\label{problemSetup}

To demonstrate the effectiveness of our proposed techniques we make
use of PCAP measurements from the Lincoln Labs DARPA Intrusion
Detection Data Set
(\url{https://www.ll.mit.edu/ideval/data/2000/LLS_DDOS_1.0.html}).
This data set cover a swath of normal computer network operations and
a variety of different attack scenarios including
\begin{itemize}
\item IP sweeps,
\item probing and breaking in through the Sadmind daemon, and
\item the preparation and execution of a Distributed Denial of Service (DDoS) attack.
\end{itemize}
We chose to use the PCAP data in our experiments since it represents
the most fundamental measurement for network problems.  However, in
future work, there are many other paths that we could consider, such as
stream based measurements.

As a preprocessing step we leveraged Wireshark
\cite{Banerjee2010,Alder2007,Sanders2011} to convert the original PCAP
files into comma separated value (CSV) files, and we then performed
the bulk of the processing using the Python \cite{Python2013}
scripting language.  The raw CSV files from Wireshark include the
following features directly extracted from the PCAP files: Source IP
address, Destination IP address, Source Port, Destination Port,
Protocol, packet length, and the packet time.

From this raw PCAP data we have chosen to create a number of higher level
features derived from this base feature set.  Perhaps most
importantly, we \emph{one-hot encode}
\cite{HastieTrevorTibshiraniRobertFriedman2009} our non-numerical
values, such as IP addresses, to create numerical values appropriate
for analysis.  For example, for each unique IP address that appears as
a source IP address in a packet header, we create a row of $Y$ which
is $1$ if that packet in from that IP address, and is $0$ otherwise.
Similarly, while a port number is an integer, port numbers do not have
the same semantics as integers (e.g., port numbers 79, the Finger
protocol, and 80, the HTTP protocol, are not really that close to each
other in functionality). Accordingly, we have rows of $Y$
corresponding to several ports that we consider to be important,
encoded using the same $0$ or $1$ scheme.

In summary, our higher level features are shown in Table~\ref{features}

\begin{table}[h]
\centering
\begin{tabular}{|p{0.9in}|p{1.9in}|}
\hline
 27 0-1 features & unique source IP addresses \\ \hline
 27 0-1 features & unique destination IP addresses \\ \hline
 13 0-1 features & important system ports on the source side \\ \hline
 13 0-1 features & important system ports on the destination side \\ \hline
 2 0-1 features & distinguish ports below 1024 from those above 1024 on the
                  source side \\ \hline
 2 0-1 features & distinguish ports below 1024 from those above 1024 on the
                  destination side \\ \hline
 1 0-1 feature & designate missing port number on the source side \\ \hline
 1 0-1 feature & designate missing port number on the destination side \\  \hline
 7 0-1 feature & various protocols
                 (ICMP, sadmind, Portmap, TELNET, TCP, FTP,
                 and HTTP) \\ \hline
 1 numerical feature & number of bytes in the packet\\ \hline
\end{tabular}
\caption{The features that we derive from PCAP data.}
\label{features}
\end{table}

This gives a $Y$ matrix with a total of 94 rows, 54 for IP addresses,
32 for ports, 7 for protocols, and 1 for packet length.

\begin{figure}[htbp]
  \centering
  \includegraphics[width=0.45\textwidth]{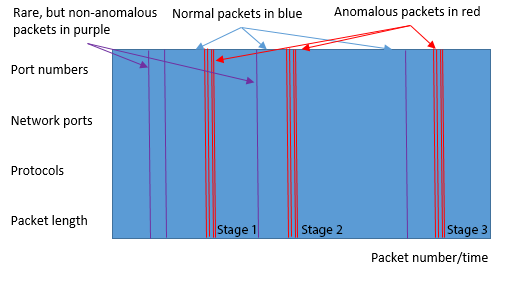}
  \caption[Caption.]{This figure shows our experimental setup. Nominal
    packets (shown in blue) are interspersed with nominal, but rare,
    events (shown in purple) and the three attack scenarios, from
    Table~\ref{scenarios} (shown in red).}
 \label{experimentalSetup}
\end{figure}

With the above features in mind, we extract from the full Lincoln Lab dataset a
subset comprised of $8322$ packets.  We then have a data matrix $Y \in
\mathbb{R}^{m \times n}$ where $m=94$ and $n=8322$.  In particular, as shown in
Figure~\ref{experimentalSetup}, our data can be viewed as a set of three attack
stages, with a large region of nominal data proceeding the first attack stage,
and smaller regions of nominal data between each of the following attack stages.
The attack stages are shown in Table~\ref{scenarios}. However, it is important
to emphasize that even the areas of the data where the attacks are occurring,
the attack packets are quite sparse and interspersed/masked by a substantial
percentage of normal data.  A cartoon of the set-up of $Y$ can be found in
Figure~\ref{experimentalSetup}.

\subsection{Decision algorithm}
\label{decisionalgorithm}

With the details from Sections~\ref{problemSetup} in mind, we can now
precisely state our method for detecting anomalies in the Lincoln Labs
DARPA Intrusion Detection Data Set.  The training of our algorithm can
be best explained by referring to Figure~\ref{experimentalSetup}.  In
particular, we take the data before the stage one attack in
Figure~\ref{experimentalSetup} and denote it by $Y_0$.  We then use
$Y_0$ to compute a nominal $L_0$ and $S_0$ using \eqref{npalg} and
\eqref{mainalg}.\footnote{Note, while this process might appear
  superficially similar to what one might do in a standard PCA
  analysis, it is actually quite different in spirit and in practice}
In particular, there are rare, but not anomalous events even before
the stage one attack that appear in $Y_0$.  Accordingly, the $S_0$
matrix is certainly not empty, even before the first attack occurs.
The nominal $L_0$ matrix is, we believe, a more accurate
representation of the true nominal processing of the network.

In effect, our goal is to identify points which are at distance greater than a
threshold $\alpha$ from the subspace spanned by $L_0$ (where distance is
measured using the $\| \cdot \|_{\infty}$ norm). Accordingly, given a training
$Y_0$ as constructed in Section~\ref{problemSetup}, and appropriate parameters
$\lambda$ and $\alpha$ (whose selection we will detail in
Section~\ref{algorithmTraining}), we can run the RPCA procedure in
\eqref{mainalg} to produce a nominal $S_0$ and $L_0$. We then extract a data
matrix $Y_a$ that overlaps one of the attack stages, but does not necessarily
overlap $Y_0$, and project $Y_a$ onto the subspace spanned by $L_0$ to compute
$L_a$.  This can be thought of as extracting the nominal part of $Y_a$.  We can
compute the anomalies for $Y_a$ by simply setting $S_a = Y_a - L_a$.  With $S_a$
in hand, our detection scheme is quite simple. We will merely flag as anomalous
any columns in $S_a$ (which is equivalent to flagging a particular packet),
whose maximum entry is larger than some threshold $\alpha$. In other words, we
flag as anomalous any packet whose corresponding column of $S_i \in S$ has $\|
S_i \|_{\infty} = \max_j | S_{ij} | > \alpha$.

\subsection{$\lambda$ and $\alpha$ training}
\label{algorithmTraining}

The selection of our values for $\lambda$ and $\alpha$ can again be best
explained by referring to Figure~\ref{experimentalSetup}.  In
particular, as before, we chose to compute our nominal $L_0$ and $S_0$
on $Y_0$.

Using our notation from Section~\ref{approach}, our first stage of
training is to compute a sequence of $L_0$ and $S_0$ matrices as we
vary $\lambda$, with small $\lambda$ values leading to lower
dimensional $L_0$ matrices and higher values of $\lambda$ leading to
higher dimensional $L_0$ matrices.  We then select the value of
$\lambda$ that leads to the best detection performance on the \emph{stage
one and stage two attacks}.  However, the $\lambda$ value we use does
not use any training data from the stage three attacks.
We then chose $\alpha$ by fixing $\lambda$ (and thereby fixing $L_0$
and $S_0$) and selecting which $\alpha$ value gives the best detection
performance over the same two stages, again not using any data from
the quite different stage three attack.
\emph{The stage three attack remains pristine}, allowing for a true
cross validation experiment
which would be similar to how the algorithm would be used in the
field.


%% file: results.tex

In this section we show a variety of results arising from our
analysis.  In particular, we demonstrate that a RPCA method, with a
$\lambda$ value \emph{optimized for attack detection} can provide
superior results to both a standard PCA analysis and a RPCA analysis
that uses a $\lambda$ value based upon some matrix recovery principal.

\begin{figure}[htbp]
  \centering
  \includegraphics[width=0.45\textwidth]{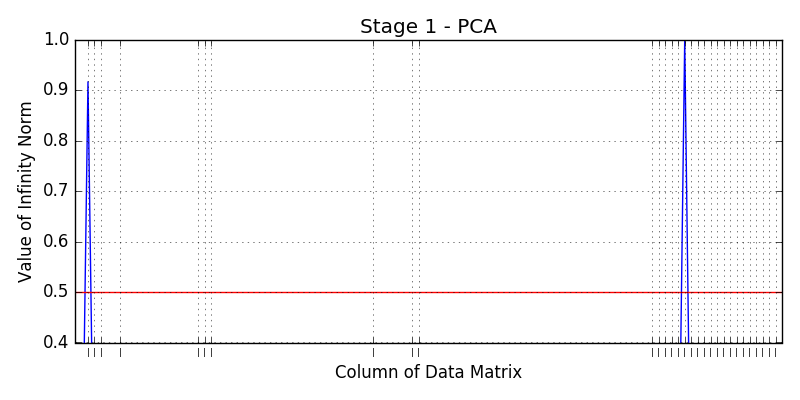}
  \includegraphics[width=0.45\textwidth]{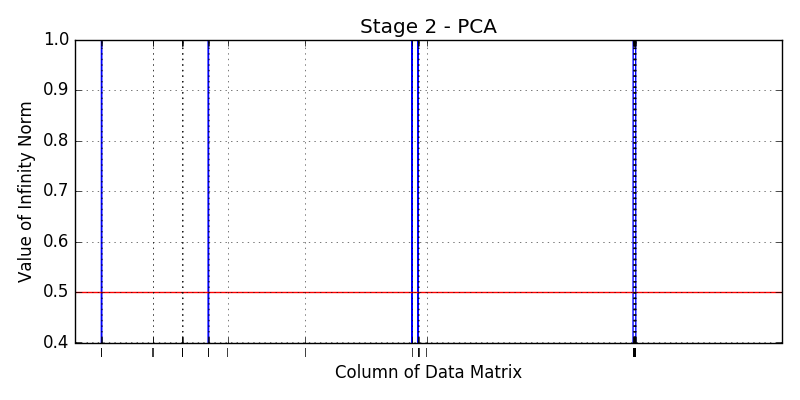}
  \includegraphics[width=0.45\textwidth]{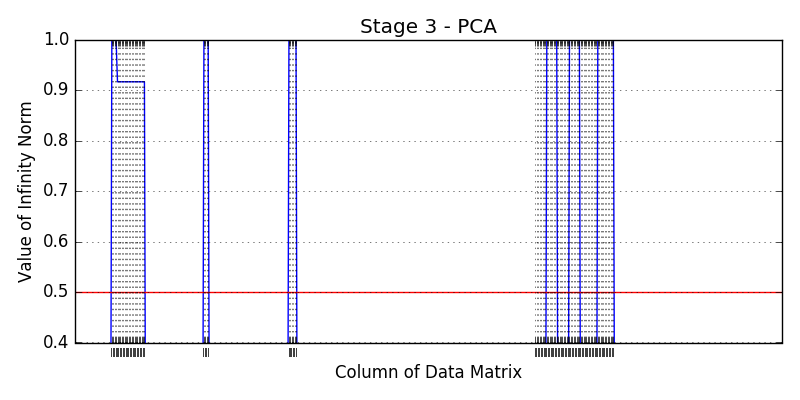}
  \caption[Caption.]{Here we show detection results for our three
    attack stages and detected using a PCA based algorithm for
    generating a low-dimensional subspace from our training data $Y_0$
    and the detection scheme discussed in
    Section~\ref{decisionalgorithm}.  The red lines represent the
    detection threshold, the blue lines show the detected anomalies
    (which raise an alarm when the blue line is above the red
    detection threshold), and black dotted lines which mark the true
    attacks.  As you can see, the PCA algorithm misses most of the
    attacks in stage one and two, as well as many of the attack packets in
    the later parts of stage three.}
\label{PCAresults}
\end{figure}

We begin with Figure~\ref{PCAresults}, where we show a standard PCA analysis. As
can be seen, the PCA analysis misses the vast majority of attack packets in all
three attack stages.  PCA is able to detect a few attack packets, but only at
the cost of numerous false negatives. We conjecture, as demonstrated in
Figure~\ref{singularvalues}, that the dimension of the subspace generated by PCA
is too large, and therefore many packets which are in fact anomalous actually
end up laying close to the normal subspace $L$ generated by the training data
$Y_0$.  It is important to note that, as already discussed in
Section~\ref{PCAsection}, that the dimension $k$ in the standard PCA algorithm
can be selected to  make the dimension of $L$ arbitrarily large or small.
However, by the very nature of the optimization \eqref{PCAopt}, the PCA
algorithm forces the detected anomalies to be \emph{small numbers}, and thereby
substantially complicates the decision algorithm in
Section~\ref{decisionalgorithm}. It is an interesting path for future research to
compare decision algorithms customized to different dimension reduction
procedures.

\begin{figure}[htbp]
  \centering
  \includegraphics[width=0.45\textwidth]{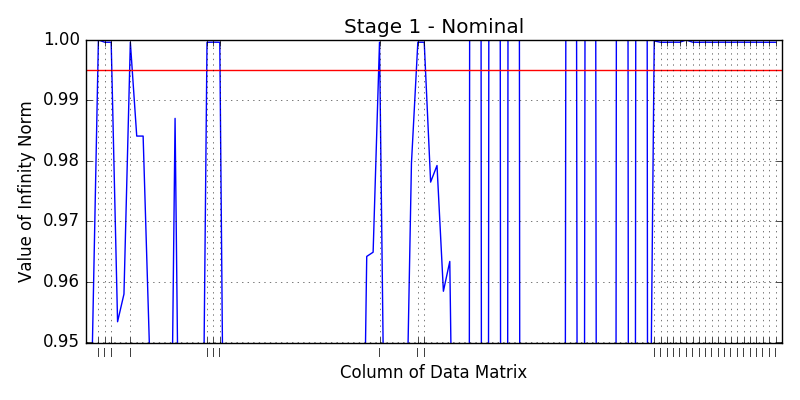}
  \includegraphics[width=0.45\textwidth]{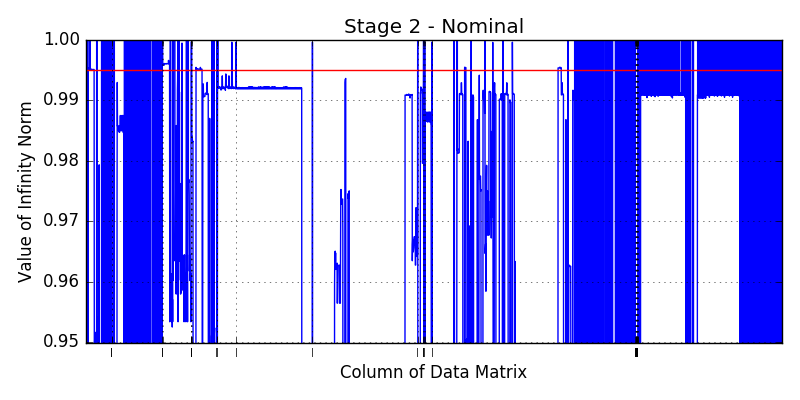}
  \includegraphics[width=0.45\textwidth]{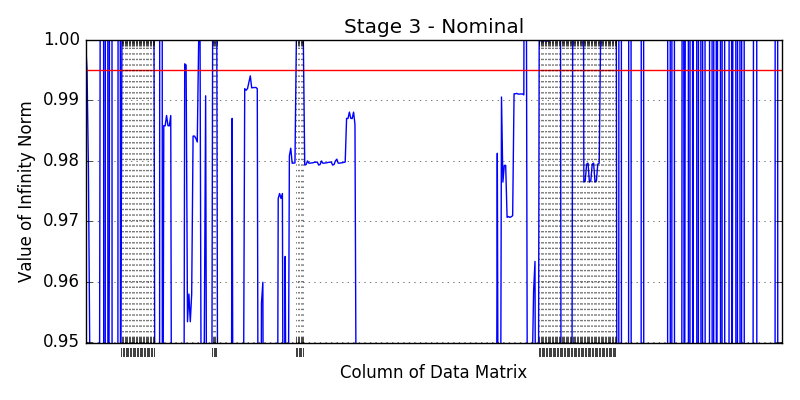}
  \caption[Caption.]{Here we show detection results for our three
    attack stages and detected using a RPCA based algorithm with a
    nominal $\lambda$ value as suggested in \cite{Candes2011}.  As in
    Figure~\ref{PCAresults}, the red lines represent the detection
    threshold, the blue lines show the detected anomalies (which raise
    an alarm when the blue line is above the red detection threshold),
    and black dotted lines which mark the true attacks.  As you can
    see, the RPCA algorithm with a nominal $\lambda$ value flags many
    normal packets as anomalous since the subspace it computed from
    the training data $Y_0$ is too low-dimensional.}
\label{nominalLambdaesults}
\end{figure}

Next, in Figure~\ref{nominalLambdaesults}, we show a RPCA analysis using a
nominal $\lambda = \frac{1}{\sqrt{\max(m,n)}} = 0.01096$, as suggested in
\cite{Candes2011} for balancing $L$ and $S$. Such a value for $\lambda$ is
classically chosen based up a recovery principle for $L$ and $S$.  As opposed to
the PCA analysis which has a large \emph{false negative} rate, the RPCA analysis
with the nominal $\lambda$ value has a large \emph{false positive} rate.  Again,
as was foreshadowed in Figure~\ref{singularvalues}, the dimension of the space
computed by RPCA from the training data $Y_0$ is much too small.  The small
nominal $\lambda$ leads to too many measurements being placed into the sparse
matrix $S$ and therefore too many detected anomalies.

\begin{figure}[htbp]
  \centering
  \includegraphics[width=0.45\textwidth]{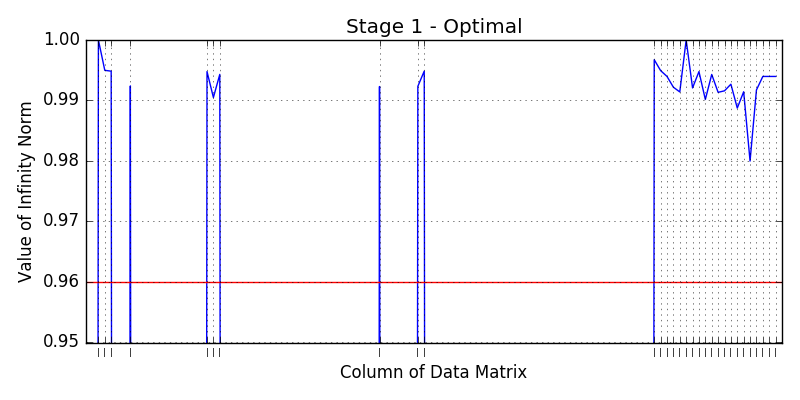}
  \includegraphics[width=0.45\textwidth]{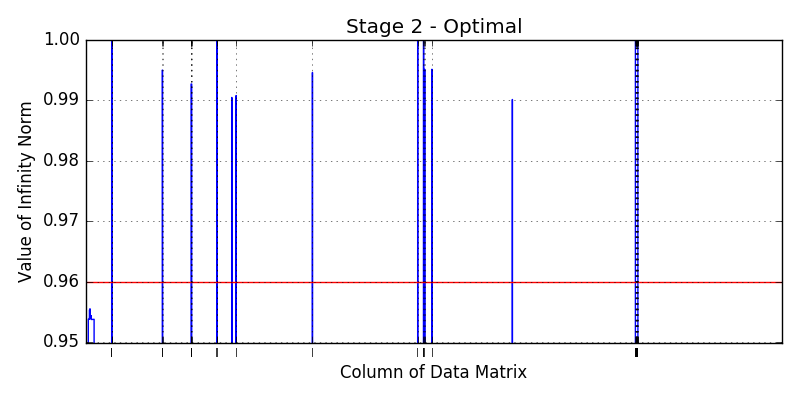}
  \includegraphics[width=0.45\textwidth]{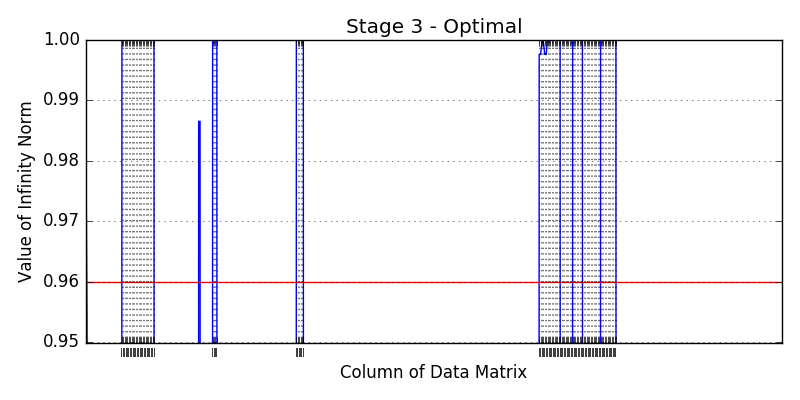}
  \caption[Caption.]{Here we show detection results for our three
    attack stages and detected using a RPCA based algorithm with an
    optimized value for $\lambda=0.157$.  As in
    Figure~\ref{PCAresults} the red lines represent the detection
    threshold, the blue line shows the detected anomalies (which raise
    an alarm when the blue line is above the red detection threshold),
    and black dotted lines which mark the true attacks.  This
    $\lambda$ value was optimized by training on stage one and two, so perhaps
    its efficacy in those stages is not overly surprising.  However,
    $\lambda=0.157$ was not computed any information from stage
    three, yet using the same subspace computed from $Y_0$ (and $\lambda$
    computed from stages one and two) gives rise to far fewer false positives
    and false negatives than the competing methods.}
\label{optimalLambdaesults}
\end{figure}

Most importantly, in Figure~\ref{optimalLambdaesults}, we show a RPCA analysis
using a $\lambda = 0.157$ which is approximately $15$ times larger
than the nominal $\lambda = 0.01096$ as in \cite{Candes2011}.
\emph{Of course, nothing in this text should be viewed as
  contradicting the results in \cite{Candes2011}.}  However, we are
solving a different problem.  We are not focused on recovering a true
low-rank $L$ and sparse $S$.  In particular, it is not even clear what
such a ``true'' low-rank $L$ and sparse $S$ would even mean in our
case.  Rather, we choose $\lambda$ to optimize our anomaly detection
performance.  In some sense, the $L$ we compute from our training data
$Y_0$ is \emph{most appropriate for the computer network at hand} and
therefore our anomaly detection performance is improved.

\begin{figure}[htbp]
  \centering
  \subfigure[\label{testa}Stage 1 ROC curve]{\includegraphics[width=0.26\textwidth]
    {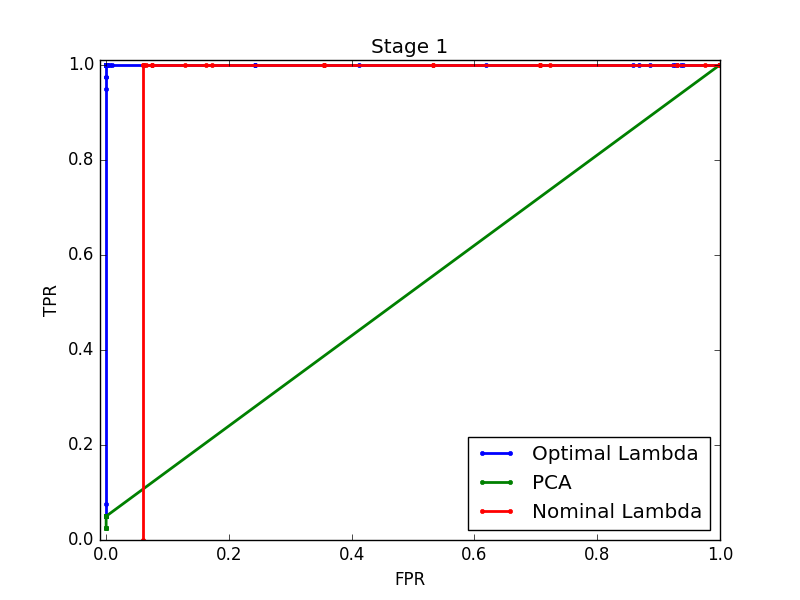}}\\
  \subfigure[\label{testb}Stage 2 ROC curve]{\includegraphics[width=0.26\textwidth]
    {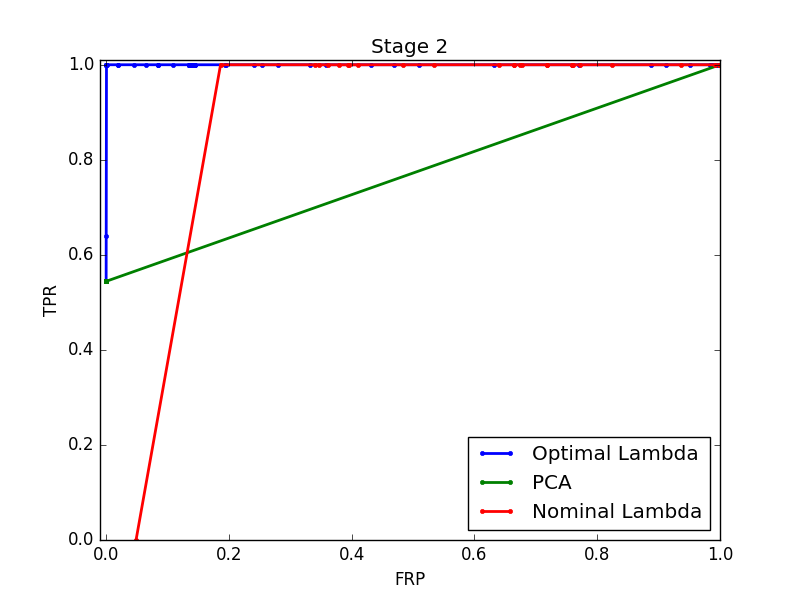}}\\
  \subfigure[\label{testc}Stage 3 ROC curve]{\includegraphics[width=0.26\textwidth]
    {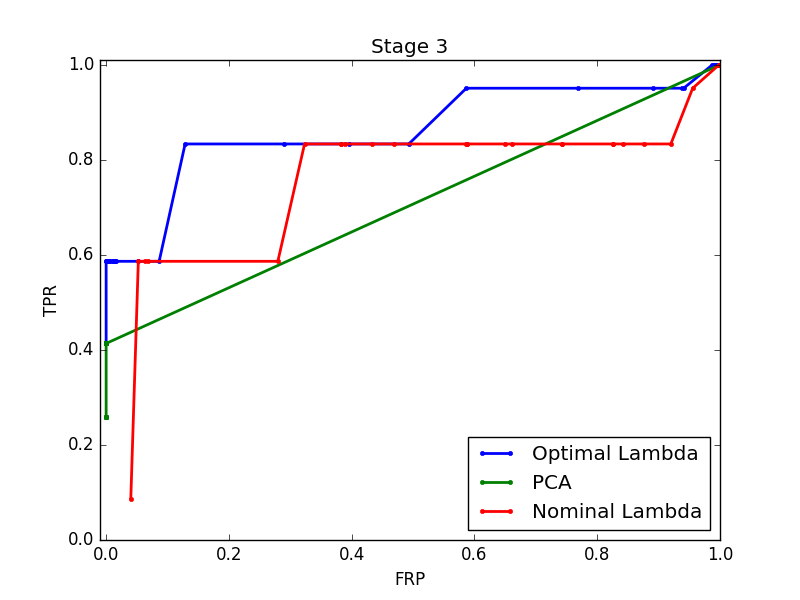}}
  \caption[Caption.]{ Here we show ROC curves, parameterized by our
    detection threshold $\alpha$, for our three attack stages.  For
    each attack stage we show the performance of PCA (shown in green),
    a nominal RPCA (shown in red), and an RPCA with a $\lambda$ value
    optimized on stage one and two (shown in blue).  We train our
    optimal $\lambda$ value on the stage one and two attacks, so
    perhaps it is not surprising that the ROC for stage one and two in
    \subref{testa} shows superior performance for the optimized RPCA
    algorithm.  However, the attack in stage 3 (in \subref{testc}) is
    \emph{quite different from the attack in stage 1}.  However, the
    ROC curves still demonstrate better performance for the trained
    $\lambda$ value, especially over a standard PCA approach.}

  \label{testFigure}
\end{figure}

Finally, as observed in Section~\ref{decisionalgorithm} we also need
to train our anomaly detection threshold $\alpha$.  Accordingly, in
Figure~\ref{testFigure} we show Receiver Operator Curves (ROC) for
each of the stages as we vary the detection threshold $\alpha$ from
$0$ to $1$.  We show ROC curves for PCA, for RPCA using a nominal
$\lambda = 0.01096$, and for RPCA using an optimized $\lambda = 0.157$.
Again, the optimized $\lambda$ was chosen by training on the attacks
in stage one and two, and therefore the good performance may not be surprising
in that case.  However, the results in stage three are \emph{not
  tuned for those specific attacks}, and the optimized $\lambda$ still
provides better performance in those cases across all threshold values
$\alpha$.

%% file: conclusions.tex
In this paper, we have demonstrated how a RPCA approach can be used to detect
anomalies in PCAP data.  In particular, we have shown that using training data
to optimize just two parameters in the RPCA algorithm, can lead to substantially
improved detection results in contrast to a PCA approach or a RPCA approach
using the literature-recommended value for $\lambda$.  We illustrated by example
that one class of attacks could quite successfully detect a separate class of
attacks.  This result supports our original hypothesis that the low dimensional
subspace computed by RPCA, even on training data, is more representative of the
true nominal state of the measured data.  This allows for a great range of
anomalies, and hence network attacks, to be successfully detected.
